\documentclass[12pt,fleqn]{article}
\usepackage{amsfonts}

\def\eq#1\en{\begin{equation}#1\end{equation}}
\def\eqa#1\ena{\begin{eqnarray}#1\end{eqnarray}}
\def\junc#1\junc{} 

\newlength{\vscaling} \newlength{\hscaling}
  \newcommand*\ti[5]{``{#5},'' {#1} {\bf #2}, #3 (#4)}

\def\Ax{\mathcal{A}_x} 

\newcommand*\scom[2]{[  #1 \, \stackrel{\star}{,} \, #2  ]}
\newcommand*\gcom[2]{[ #1 \, , \, #2  ]_\mathrm{G}}
\newcommand*\sncom[2]{[  #1 \, , \, #2 ]_\mathrm{S}}
\newcommand*\pcom[2]{\{ #1 \, , \, #2 \}}

\newcommand*\Hom{\mathrm{Hom}}

\newcommand*\dpo[1][]{\mathbf{d}_{\theta #1}} 
\newcommand*\ds[1][]{\mathbf{d}_{\star #1}}   
\newcommand*\dst{\mathbf{\tilde d}_{\star}}    

\newcommand*\ap[1][]{\mathbf{a}_{\theta #1}} 
\newcommand*\as[1][]{\mathbf{a}_{\star #1}}   
\newcommand*\fp[1][]{\mathbf{f}_{\theta #1}} 
\newcommand*\fs[1][]{\mathbf{f}_{\star #1}}   

\newcommand*\As[1][]{\mathcal{A}_{#1}} 

\newcommand*\ha{\hat A}
\newcommand*\hl{\hat \lambda}
\newcommand*\hd{\hat\delta}

\def\id{\mathrm{id}}

\def\pp{\partial}

\def\EE{{\cal E}}
\def\AA{{\cal A}}
\def\FF{{\cal F}}

\def\MM{{\cal M}}

\def\DD{{\cal D}}
\def\OO{{\cal O}}
\def\CC{{\cal C}}

\begin{document}
\begin{titlepage}
\rightline{LMU-TPW 00-25}
\vfill
\begin{center}
{\bf\LARGE
Nonabelian noncommutative gauge fields and Seiberg-Witten map}

\vfill

{{\bf Branislav Jur\v co, Peter Schupp and Julius Wess
}}

 \vskip 0.5 cm

Universit\"at M\"unchen,
Theoretical Physics Group\\ Theresienstr.\ 37,
80333 M\"unchen, Germany
\end{center}
\vfill
\begin{abstract} 
Noncommutative gauge
fields (similar to the type that arises in string theory with background $B$-fields)
are constructed for arbitrary nonabelian gauge groups with the help of a map that 
relates ordinary nonabelian  and noncommutative gauge theories
(Seiberg-Witten map). 
As in our previous work we employ Kontsevich's formality 
and the concept of equivalent star products. As a byproduct we obtain a
``Mini Seiberg-Witten map'' that explicitly relates ordinary abelian and
nonabelian gauge fields.
(This paper is based on a talk by P.~Schupp at the ``Brane New World''
conference in Torino; for a more detailed version see~\cite{JSW}.)
\end{abstract}
\vfill
\hrule
\vskip 5pt
\noindent
{\footnotesize\it e-mails: \parbox[t]{.8\textwidth}{%
jurco\,,\,schupp\,,\,wess\,@\,theorie.physik.uni-muenchen.de}}
\end{titlepage}\vskip.2cm

\newpage

\setcounter{page}{1}

\section{Introduction}

The topic of this talk is the type of noncommutative gauge theory
that arises in string theory~\cite{Seiberg:1999vs}.
Since it is not yet clear what exactly the relevant ingredients of
the final form of the theory will be we shall not use an axiomatic
approach but will instead try to motivate everything from physics
as we go along.
We choose to work in the framework of deformation quantization because
it allows a mathematically rigorous formulation and solution of the
questions that interest us here.%
\footnote{Clearly one has to be careful with star products -- in particular
when studying gauge Morita equivalence of noncommutative tori, where
the deformation parameter could be mapped to its inverse~\cite{Connes:1998cr}. 
Nevertheless
for the construction of the Seiberg-Witten map it appears to be
the right language; if there is a solution of the recursion relations
that define the map then it should have the form presented here.}

Let us start by briefly recall how star products and
noncommutative gauge theory arises
in string theory:
Consider an open string $\sigma$-model with background
term
\eq
S_B = \frac{1}{2i} \int_D B_{i\!j}\, \epsilon^{ab} \pp_a x^i \pp_b x^j ,
\en
where the integral is over the string world-sheet (disk) and
$B$ is constant, nondegenerate and $dB = 0$. The correlation functions
on the boundary of the disc in the decoupling limit ($g \rightarrow 0$,
$\alpha' \rightarrow 0$) are
\eq
\left\langle f_1(x(\tau_1)) \cdot\ldots\cdot f_n(x(\tau_n))\right\rangle_B
= \int dx \, f_1 \star \ldots
\star f_n , \quad (\tau_1 < \ldots < \tau_n) \label{corr}
\en
with the Weyl-Moyal star product
\eq
(f \star g ) (x)  = \left. e^{\frac{i\hbar}{2}\theta^{i\!j} \pp_i \pp_j'} f(x) g(x') \right|_{x'
\rightarrow x} ,
\en
which is the deformation quantization of the Poisson structure $\theta =
B^{-1}$.
More generally a star product is an associative, $[[\hbar]]$-bilinear product
\eq
f \star g = f g + \sum_{n=1}^\infty (i\hbar)^n
\underbrace{B_n(f,g)}_{\mathrm{bilinear}} ,
\en
which is the deformation of a Poisson structure $\theta$:
\eq
\scom{f}{g} = i \hbar \pcom{f}{g} + \OO(\hbar^2), \quad \pcom{f}{g} =
\theta^{i\!j}(x) \pp_i f \, \pp_j g .
\en
Let us now perturb the constant $B$ field by adding a gauge potential $a_i(x)$:
$B \rightarrow B + da$, $S_B \rightarrow S_B + S_a$, with
\eq
S_a = -i \int_{\partial D} d\tau a_i(x(\tau))\partial_\tau x^i(\tau).
\en
Classically we have the naive gauge invariance
\eq
\delta a_i = \partial_i \lambda, \label{naive}
\en
but in the quantum theory this depends on the choice of regularization.
For Pauli-Villars (\ref{naive}) remains a symmetry but
if one expands $\exp S_a$ and employes a point-splitting regularization
then the functional integral is invariant under noncommutative gauge
transformations\footnote{This formula is only valid for the Moyal-Weyl
star product.}
\eq
\hat\delta \ha_i = \partial_i \hl + i \hl \star \ha_i - i \ha_i \star \hl.
\label{nctrans}
\en
Since a senisible quantum theory should be independent of the choice of regularization
there should be field redefinitions 
$\ha(a)$, $\hl(a,\lambda)$ (Seiberg-Witten map) that relate (\ref{naive}) and
(\ref{nctrans}):
\eq
\ha(a) +\hat\delta_{\hl} \ha(a) = \ha(a+\delta_\lambda a). \label{swcond}
\en
The problems that we shall address are:
\begin{itemize}
\item A direct construction of
maps $\ha(a)$, $\hl(a,\lambda)$ to all orders in the deformation
parameter that satisfy
the compatibility condition (\ref{swcond}); not term by term via a
recursion relation.
\item This construction should work for any
$\star$-product, i.e., any Poisson structure $\theta(x)$ and it should
be covariant under general coordinate transformations. (For this
we first need to give a suitably general formulation of noncommutative
gauge theory.)
\item A further generalization to arbitrary
nonabelian gauge groups; not just by simply absorbing a matrix factor into
the definition of the algebra.
\end{itemize}
Sections~\ref{covcord} and \ref{swmap} review the
results of~\cite{Madore:2000en}, \cite{Jurco:2000fb}, \cite{Jurco:2000fs}
in a slightly more general setting. Section~\ref{nonabelian} discusses 
nonabelian gauge fields \cite{Bonora:2000td},
\cite{Jurco:2000ja} in this framework.

In related work a path integral approach 
\cite{Ishibashi:1999vi}, \cite{Okuyama:2000ig} and operator
methods~\cite{Cornalba:2000ua} have been discussed;
the role of  $\star_n$-operations has been illuminated
in~\cite{Mehen:2000vs}, \cite{Liu:2000mj}.

\section{Noncommutative gauge fields}

To see how to proceed we note that the extra factor $\exp S_a$ in
the correlation function (\ref{corr}) effectively shifts the
coordinates\footnote{Notation: $\DD$ should not be confused with a covariant
derivative (but it is related).}
\eq
x^i \rightarrow x^i + \theta^{ij}\ha_j =: \DD x^i . \label{covco}
\en
More generally, for a function $f$,
\eq
f \rightarrow f + \AA(f) =: \DD f. \label{covfu}
\en
$\AA$ plays the role of a generalized gauge potential; it maps a function
to a new function that depends on the gauge potential.
The shifted coordinates and functions are covariant under noncommutative
gauge transformations:
\eq
\hat\delta (\DD x^i) = i\scom{\hl}{\DD x^i}, \qquad
\hat\delta (\DD f) = i\scom{\hl}{\DD f}.
\en
The first expression implies (\ref{nctrans}) (for $\theta$ constant and
nondegenerate).

\subsection{Covariant coordinates, covariant functions}
\label{covcord}

The covariant coordinates (\ref{covco}) are the
background independent operators of \cite{Seiberg:1999vs,Seiberg:2000zk}; 
they and the covariant functions (\ref{covfu}) can also
be introduced more abstractly as follows:
Let $\Ax$ be an associative algebra ``noncommutative space''
with product $\star$.\footnote{Until section~\ref{swmap}
this does not need to be a star product} In reference to their
commutative analog we shall call the elements of $\Ax$ functions.
The gauge transformation of a field $\psi \in$ (left module of) $\Ax$
is
\eq
\hat\delta \psi = i \hl \star \psi, \label{deltapsi}
\en
where the gauge parameter $\hl \in \Ax$ is an arbitrary function on
the noncommutative space. (More precisely $\psi$ is the coefficient
function (or vector) of a section in a projective module over $\Ax$
-- we shall come back to this later.)
Since the product of a function and a field is not covariant on a noncommutative
space,
\eq
\hat\delta(f\star\psi) = f\star\hd\psi = f\star(i\hl\star\psi)
\neq i \hl \star(f\star\psi),
\en
one needs to introduce covariant functions
\eq
\DD f = f + \AA(f),
\en
where $\AA \in \Hom(\Ax,\Ax)$ transforms such that
\eq
\hd (\DD f \star\psi) = i\lambda \star (\DD f \star \psi),
\en
i.e.,
\eq
\hd \AA(f) = \scom{i\hl}{f} + \scom{i\hl}{\AA(f)}.
\en
There are other covariant objects:
\eq
\FF(f,g) = \scom{\DD f}{\DD g} - \DD(\scom{f}{g}), \label{field}
\en
e.g., plays the role of a generalized field strength; it maps two functions to
a new function that depends on the gauge potential and transforms covariantly.
The generalized field strength is antisymmetric in its arguments, i.e.,
$\FF \in \Hom(\Ax^{\wedge 2},\Ax)$.
For $\theta$ constant $\AA(x^i) = \theta^{ij} \hat A_j$ and
\eq
\FF(x^i,x^j) = i\theta^{ik} \theta^{jl} \hat F_{kl},
\qquad
\hat F_{kl} = \partial_k\ha_l -\partial_l\ha_k - i\scom{\ha_k}{\ha_l}.
\en
There are several reasons, why one needs $\AA$ and $\DD$ and not just
$A^i \equiv \AA(x^i)$ (or $\hat A_i$, for $\theta$ constant):
If we perform a general coordinate transformation
$x^i \mapsto {x^i}'(x^j)$ and
transform $A^i$ (or $\hat A_i$) as its index structure suggests,
then we will obtain objects that are no longer covariant under
noncommutative gauge transformations. The correct transformation
is $\AA(x^i) \mapsto \AA({x^i}')$. Furthermore we may be interested
in covariant versions of scalar fields $\phi(x)$. These are given
by the corresponding covariant function $\DD(\phi(x))$.

\subsection{Abstract formulation of gauge theory on a noncommutative
space}

Finite projective modules take the place of fiber bundles in the noncommutative
realm. This is also the case here, but may not have been apparent 
since we have been working with component
fields as is customary in the physics literature.
As we have argued, $\AA \in C^1$, $\FF \in C^2$ with
$C^p = \Hom(\Ax^{\wedge p},\Ax)$, $C^0 \equiv \Ax$. These $p$-cochains take the place of forms
on a noncommutative space, the wedge product is replaced by the cup
product $\wedge: C^p \otimes_{\Ax} C^q \rightarrow C^{p+q}$ and the exterior
differential is replaced by the standard coboundary operator $\ds$
in the Lie-algebra cohomology;  $\ds^2 = 0$, $\ds 1 = 0$.
(In the Hochschild cohomology $\ds$ can be expressed in terms of the Gerstenhaber
bracket, $\ds\CC = -\gcom{\CC}{\star}$, and can then be restricted to a map
$C^p \rightarrow C^{p+1}$
by antisymmetrization.)
This calculus  uses only the algebraic structure of $\Ax$; it is
related to the standard universal calculus and  we can obtain
other calculi by projection.
Consider now a (finite) projective right $\Ax$-module $\EE$.
We can introduce a connection on $\EE$ as a linear map
$\nabla : \EE \otimes_{\Ax} \!C^p \rightarrow \EE \otimes_{\Ax} C^{p+1}$
for $p \in \mathbb{N}_0$
which satisfies the Leibniz rule\footnote{The transformation of matter
fields (\ref{deltapsi}) leads to a slight complication here; for fields
that transform in the adjoint (by star-commutator) we only need $\tilde\nabla$,
$\ds$.} 
\eq
\nabla(\eta \psi)
 =  (\tilde\nabla \eta) \psi
+ (-)^p \eta\,\dst\psi \nonumber
\en
for all $\eta\in\EE \otimes_{\Ax} \!C^p$, $\psi \in C^r$, and where
$\tilde\nabla\eta = \nabla\eta - (-)^p \eta\,\dst 1$,
\eq
\dst(a\wedge\psi) = (\ds a)\wedge\psi + (-)^q a \wedge(\dst\psi)
\en
for all $a \in C^q$, and $\dst 1$ is the identity operator on $\Ax$.

Let $(\eta_a)$ be a  generating family for $\EE$;
any $\xi$ can then be written as $\xi = \sum \eta_a \psi^a$ with $\psi^a \in \Ax$
(with only a finite number of terms different from zero). For a free module
the $\psi^a$ are unique, but we shall not assume that. Let the generalized
gauge potential be defined by the action of $\tilde\nabla$ on the elements
of the generating family: $\tilde\nabla \eta_a = \eta_b \AA^b_a$. In the
following we shall suppress indices and simply write $\xi = \eta.\psi$,
$\tilde\nabla \eta = \eta.\AA$ etc. We compute
\eq
\nabla\xi = \nabla(\eta.\psi) = \eta.(\AA\wedge\psi + \dst \psi)
= \eta.(\DD\wedge\psi).
\en
Evaluated on a function $f\in\Ax$ this yields a covariant function
times the matter field: $(\DD\wedge\psi) (f) = (f + \AA(f))\star\psi
=(\DD f)\star\psi$.
Similarily 
\eq
\nabla^2 \xi = \eta.(\AA\wedge\AA + \ds\AA).\psi = \eta.\FF.\psi
\en
with the field strength
\eq
\FF =  \ds\AA + \AA \wedge \AA,
\en
which  agrees with (\ref{field}) and satisfies the Bianchi identity
\eq
\ds\FF + \AA\wedge\FF - \FF\wedge\AA = 0
\en
due to the associativity of $\Ax$.
Infinitesimal gauge transformations in the present notation are
\eq
\delta\AA = -i \ds\lambda + i\lambda\wedge\AA - i\AA\wedge\lambda,
\en
\eq
\delta\FF =  i\lambda\wedge\FF - i\FF\wedge\lambda.
\en

\section{Equivalence of star products and Seiberg-Witten map}
\label{swmap}

Reconsider the correlation function (\ref{corr}) from the point
of view of the
path-integral representation~\cite{CattaneoFelder} of 
Kontsevich's star product.
This is not limited to constant Poisson structures, so we can
study the result of the perturbation $B \rightarrow B + da$
more directly:\footnote{%
Here we have $g \equiv 0$ from
the start and Batalin-Vilkovisky quantization has to be used,
while in (\ref{corr}) $g$ was send to zero after computing the correlation
function.}
\[
\langle f(x(0)) g(x(1)) \delta_x(x(\infty))\rangle_B = (f\star g)(x),
\]
\eq
\langle f(x(0)) g(x(1)) \delta_x(x(\infty))\rangle_{B + da} = (f\star' g)(x).
\en
More generally we can consider Poisson structures $\theta$, $\theta'$
(which may be degenerate)
in place of the symplectic structures $\omega = B$ and $\omega' = B + da$:
\eq
\begin{array}{ccc}
   \mbox{Poisson structure} & & \mbox{star product} \\
 \theta & \longrightarrow   & \star \\
   \downarrow &&\\
 \theta' & \longrightarrow & \star'
\end{array}
\en
Interpretation and Strategy: The map from  $\star$ to $\star'$ that completes
the above diagram is the equivalence map $\DD = \id + \AA$ defined by the
generalized gauge potential $\AA$:
\eq
\DD f \star \DD g = \DD(f \star' g).
\en
In the nondegenerate case we can argue for this from Moser's 
lemma~\cite{Moser}:
The symplectic structures $\omega$ and $\omega'
= \omega + da$ are related by a coordinate transformation $\rho^*$
generated by the vector field $\theta^{ij}a_j\partial_i$:\footnote{More
precisely we should also ask that $\omega_t \equiv \omega + t\, da$ is
nondegenerate for all $t \in [0,1]$.}
\eq
\rho^*\{f,g\}' = \{\rho^* f , \rho^* g\}.
\en
It then follows from a theorem due to Kontsevich~\cite{Kontsevich:1997vb}
that the corresponding
star products $\star$ and $\star'$ are equivalent. In first order in
$\theta$ the equivalence map and $\AA$ are  given by Moser's 
vector field, i.e.,
in terms of the ordinary gauge potential~$a_i$.
In the next sections we shall construct this equivalence map $\DD$ and
the generalized noncommutative gauge potential $\AA$ explicitly as
a function of the ordinary gauge potentials $a$.

\subsection{Abelian case}

We shall take an abelian gauge potential, the corresponding field
strength and abelian gauge transformations as given data:
\eq
a = a_i dx^i , \quad f = da, \quad \delta_\lambda a = d\lambda .
\en
We will first construct a semiclassical version of the Seiberg-Witten map,
where all star commutators are replaced by Poisson brackets. The construction
is essentially a formal generalization of Moser's lemma to Poisson manifolds.

\subsubsection{Semi-classicaly}

Let us introduce the coboundary operator
\eq
\dpo = -\sncom{\,\cdot}{\theta}, \label{dtheta}
\en
where $\sncom{\,}{\,}$ is the Schouten-Nijenhuis bracket
of polyvector fields
and $\theta = \frac{1}{2} \theta^{ij} \partial_i\wedge\partial_j$ is
the Poisson bivector.
($\dpo$ is a very useful operator, since it turns a function $f$ into the
corresponding Hamiltonian vector field $\dpo f = \theta^{ij}(\partial_j f)
\partial_i \equiv \pcom{\,\cdot}{f}$.)
Let us define
\eq
\ap = a_i\, \dpo x^i = \theta^{ji} a_i \partial_j \qquad \mbox{(Moser's vector field)},
\label{vmoser}
\en
\eq
\fp = \dpo \ap =
-\frac{1}{2}(\theta\cdot f\cdot\theta)^{ij}\partial_i\wedge\partial_j 
\en
and introduce a one parameter deformation $\theta_t$ of the Poisson structure
$\theta$: $t \in [0,1]$, $\theta_0 = \theta$, $\theta_1 =: \theta'$ by the
differential equation
\eq
\partial_t \theta_t = \fp[_t]  . \label{diff}
\en
($\theta_t$ is Poisson, because $\dpo[_t]\fp[_t] = 0$ and if $f$ is not explicitely
$\theta$-dependent, we find
$\theta_t = \theta - t\,\theta f\theta + t^2\, \theta f
\theta f \theta \mp \ldots$,
i.e., $\omega_t = \omega + t f$ in the nondegenerate case.)
The $t$-evolution is generated by the vector field $\ap$,
because $\fp[_t] = -\sncom{\ap[_t]}{\theta_t}$.
The Poisson structures $\theta$ and $\theta'$ are related by the flow
\eq
\rho_a^* = \left. e^{\ap[_t] + \partial_t} e^{-\partial_t} \right|_{t=0}.
\en
Let
\eq
A_a = \rho_a^* - \id, \qquad \tilde\lambda = \sum_{n=0}^{\infty} 
\left.\frac{(\ap[_t] +
\partial_t)^n(\lambda)}{(n+1)!}\right|_{t=0},
\en
then:
\eq
A_{a+d\lambda} = A_a + \dpo \tilde\lambda + \pcom{A_a}{\tilde\lambda}.
\en
This is the semi-classical version of the Seiberg-Witten condition (\ref{swcond}).

\subsubsection{Kontsevich formality map}

We would now like to quantize the semi-classical solution by lifting
the Poisson structures to star products and, indeed, by lifting all 
polyvector fields
to polydifferential operators. This is done by Kontsevich's formality
maps~\cite{Kontsevich:1997vb}:
\eq
U_0 (f) = f, \quad
\underbrace{\big[U_n(\underbrace{\alpha_1,
\ldots,\alpha_n}_{\mbox{polyvectors}})\big]}_{\mbox{plolydiff.\
operator}}(\underbrace{f_1,
\ldots,f_m}_{\in\, C^\infty(\MM)})
\;\in\; C^\infty(\MM), \quad n \in \mathbb{N},
\en
\eq
\alpha_i = \sum[\alpha_i(x)]^{l_1 \ldots l_{k_i}} \partial_{l_1}
\wedge\ldots\wedge\partial_{l_{k_i}}.
\en
Some properties of the $U_n$:
\begin{itemize}
\item skew-symmetric, multilinear in all $n$ arguments
\item matching condition: $m = 2 - 2n + \sum k_i$
\item formality condition (FC) (see \cite{Kontsevich:1997vb},
\cite{Manchon})
\end{itemize}
Example: The $U_n$ lift a bivector field $\theta$ to the bidifferential operator
\eq
f \star g = \sum \frac{(i\hbar)^n}{n!} U_n(\theta,\ldots,\theta)(f,g) 
= f g + \frac{i\hbar}{2} \theta^{ij}\partial_i f \partial_j g + \ldots \, .
\label{star}
\en
If $\sncom{\theta}{\theta} = 0$ (Poisson) then the formality condition implies
that $\star$ is associative. For constant $\theta$
(\ref{star}) is precisely the Moyal-Weyl star product.

\subsubsection{Quantum}
\label{poisson}

We lift Moser's vector field $\ap$ via formality to the differential operator
\eq
\as = \sum \frac{(i\hbar)^n}{n!} U_{n+1}(\ap,\theta,\ldots,\theta),
\en
whose transformation properties follow from the formality condition:
\eq
\delta\ap = \dpo\lambda \;\stackrel{\mathrm{(FC)}}{\Rightarrow}\;
\delta\as = \frac{1}{i\hbar} \ds\bar\lambda; \quad
\bar\lambda \equiv \sum\frac{(i\hbar)^n}{n!}U_{n+1}(\lambda,\theta,\ldots,\theta).
\en
In analogy to (\ref{dtheta}) and (\ref{vmoser}) we have the
coboundary operator
\eq
\ds = -\gcom{\, \cdot}{\star\,} ,
\en
where $\gcom{\,}{\,}$ is the Gerstenhaber bracket, and the bidifferential operator
\eq
\fs = \ds \as \stackrel{\mathrm{(FC)}}{=}
\sum \frac{(i\hbar)^{(n+1)}}{n!} U_{n+1}(\fp,\theta,\ldots,\theta).
\en
We lift $\theta_t$ to a $t$-dependent star product
\eq
g \star_t h = \sum \frac{(i\hbar)^n}{n!} U_n(\theta_t, \ldots, \theta_t)(g,h)
\;\stackrel{\mathrm{(FC)}}{\Rightarrow}\; 
\partial_t (g \star_t h) = \fs[_t] (g,h) .
\en
As an operator equation: $\partial_t(\star_t) = \fs[_t]$.
Since $\fs[_t] = -\gcom{\as[_t]}{\star_t}$,
the $t$-evolution is generated by the differential operator $\as$.
The star products
$\star$ and $\star'$ are related by the equivalence map or ``quantum flow''
\eq
\DD_a = \left. e^{\as[_t] +\partial_t} e^{-\partial_t}\right|_{t=0} .
\en
Let
\eq
\As[a] = \DD_a - \id, \qquad \Lambda(\lambda,a) = \sum\left.\frac{(\as[_t] +
\partial_t)^n(\bar\lambda)}{(n+1)!}\right|_{t=0}, \label{sw}
\en
then
\eq
\As[a + d\lambda] = \As[a] + \frac{1}{i\hbar}(\ds \Lambda - \Lambda\star\As[a]
+ \As[a]\star\Lambda).
\en
In (\ref{sw}) we have thus found the full Seiberg-Witten map for the
generalized noncommutative gauge potential $\As$ for an arbitrary
Poisson structure $\theta(x)$. In components:
\eq
\As[a](x^i) = \theta^{ij} a_j + \frac{1}{2} \theta^{kl}a_l (\partial_k
(\theta^{ij} a_j) - \theta^{ij} f_{jk}) + \ldots ,
\en
\eq
\Lambda = \lambda + \frac{1}{2} \theta^{ij} a_j \partial_i\lambda +
\frac{1}{6} \theta^{kl} a_l (\partial_k(\theta^{ij} a_j \partial_i \lambda)
-\theta^{ij} f_{jk} \partial_i \lambda) + \ldots .
\en

\subsection{Nonabelian case}
\label{nonabelian}

The given nonabelian data is
\[
A_\mu(x) = A_{\mu b}(x) T^b, \quad [T^a,T^b] = i C^{ab}_c T^c,
\]
\eq
\delta_\Lambda A_\mu = \partial_\mu \Lambda + i[\Lambda, A_\mu], \quad
F_{\mu\nu} = \partial_\mu A_\nu - \partial_\nu A_\mu - i[A_\mu,A_\nu].
\en
Our goal is to find maps $\hat A(A_\mu)$, $\hat\Lambda(A_\mu,\Lambda)$, such
that
\eq
\hat A(A_\mu + \delta_\Lambda A_\mu) = \hat A(A_\mu) + \hat\delta_{\hat\Lambda}
\hat A(A_\mu),
\en
with a noncommutative (and nonabelian) gauge transformation $\hat\delta$.
A naive geralization of the abelian construction immediately runs into several
problems: We could try to let $B \rightarrow B + F =: B'$ and correspondingly
$\Theta \rightarrow \Theta'$, but even if $dB = 0$ ($\Theta$ Poisson)
$dB' \neq 0$ and $\Theta'$ will not be Poisson, since $dF = -A\wedge A \neq 0$.
Furthermore, the nonabelian analog of Moser's vector field,
$\Theta^{\mu\nu}A_\nu D_\mu$,
is not a derivation, essentially because $A_\nu$ is matrix-valued.

Strategy: We shall introduce a larger space with coordinates $x^i$, $x^j$,~
\ldots\ that is the product of noncommutative space-time
with coordinates $x^\mu$, $x^\nu$,~\ldots\ and an internal space with coordinates
$t^a$, $t^b$,~\ldots. The enlarged space shall
have a Poisson structure which is the direct sum of external $\Theta^{\mu\nu}$ and
internal $\vartheta^{ab} = C^{ab}_c t^c$ Poisson structures
\eq
\left(\theta^{ij}\right) = \left(\begin{array}{c|c} \Theta^{\mu\nu} & 0\\
\hline
0 & \vartheta^{ab} \end{array}\right).
\en
Then we pick appropriate \emph{abelian} gauge potentials and parameters
that are linear in the internal coordinates and quantize everything with
the previous method.
After quantization of the internal space
(and taking an appropriate representation) the symbols
$t^a$ become the generators of the Lie algebra with structure constants
$C^{ab}_c$, giving an nonabelian theory on commutative space-time.
After quantizing everything we obtain the desired nonabelian noncommutative
gauge theory.

\subsection{Mini Seiberg-Witten map}

As in the abelian case the space-time components
of the noncommutative gauge potential
are determined via the Seiberg-Witten map by the term of
lowest order in $\Theta$:
\eq
\As[a](x^\mu) = \Theta^{\mu\nu} A_\nu(a) + \OO(\Theta^2).
\en
From (\ref{sw}) we can compute the nonabelian gauge potential
$A_\nu(a)$ and the nonabelian gauge parameter $\Lambda$
to all orders in $C^{ab}_c$ and in the internal components of the
abelian gauge potential $a_b$:
\eq
A_\nu(a) = \sum \frac{1}{n!} t^c (M^{n-1})^b_c (a_{\nu b} - (n-1) f_{\nu b}),
\en
with $a_{\nu b} t^b = a_\nu$ and the matrix $M^a_b = C^{ac}_b a_c$,
\eq
\Lambda(\lambda,a) = \sum\frac{1}{(n+1)!} t^a (M^{n})^b_a \lambda_b.
\en
(In the special gauge of vanishing internal gauge potential $a_b =0$
this becomes simply $A_\mu(a) = a_\mu$, $\Lambda = \lambda$.)

\section*{Acknowledgements}

We would like to thank A.\ Alekseev, P.\ Aschieri, J.\ Madore,
S.\ Schraml and A.\ Sitarz for helpful discussions.
B.J.\ and P.S.\ thank the
LPM Universit\' e de Montpellier for hospitality and P.S.\
thanks the participants of the Torino conference, in particular
A.\ Schwarz, for interesting and clarifying discussions
and the organizers for a very nice conference.

\end{document}